\documentclass[12pt]{article}

\newcommand{\sect}[1]{\section{#1}\setcounter{equation}{0}}

\begin{document}
\baselineskip 16pt

\def\G{\Gamma}
\def\oG{|\G|}
\def\Mg{M/\G}
\def\s3g{S^3/\G}
\def\phi{\varphi}
\def\gsim{\; \raisebox{-.8ex}{$\stackrel{\textstyle >}{\sim}$}\;}
\def\lsim{\; \raisebox{-.8ex}{$\stackrel{\textstyle <}{\sim}$}\;}
\def\beq{\begin{equation}}
\def\eeq{\end{equation}}
\def\({\left (}
\def\){\right )}
\def\[{\left [}
\def\]{\right ]}
\def\T{{\bf T}}

\def\phi{\varphi}
\def\g{\gamma}
\def\a{\alpha}
\def\b{\beta}
\def\tphi{\tilde{\phi}}
\def\ra{\rangle}
\def\la{\langle}

\def\R{{\cal R}}
\def\W{{\cal W}}
\def\V{{\cal V}}

\bigskip
\hspace*{\fill}
\bigskip\bigskip\bigskip

\begin{center}
\Large \bf Note on Gauge Theories on $M/\G$  \\
and the AdS/CFT Correspondence
\end{center}
\bigskip\bigskip\bigskip
\centerline{\large Gary T. Horowitz${}^*$ and Ted Jacobson${}^\dagger$\footnote{\tt 
gary@cosmic.physics.ucsb.edu, jacobson@physics.umd.edu}}
\medskip
\centerline{${}^*$Department of Physics, UCSB, Santa Barbara, CA 93106-9530,}
\centerline{${}^\dagger$Department of Physics, University of Maryland, College Park, MD}

\begin{abstract}
It is well known that a weakly coupled $U(N)$ gauge theory on a
torus with sides of length $L$ has extra light states with
energies of order $1/(NL)$.  We show that a similar result holds
for gauge theories on $\Mg$ where $M$ is any compact Riemannian manifold
and $\G$ is any freely acting discrete isometry group.  As in the
toroidal case, this is achieved by adding a suitable nontrivial
flat connection.  As one application, we consider the AdS/CFT
correspondence on spacetimes asymptotic to $AdS_5/\G$.  By
considering finite size effects at nonzero temperature, we show
that consistency requires these extra light states of the gauge
theory on $S^3/\G$.
\end{abstract}

\newpage

\sect{Introduction}

Consider a weakly coupled 
$U(N)$
gauge theory on an $m$-dimensional torus where the length of each
circle is equal to $L$.  Naively, one might expect that the
lowest energy states will have energy of order $1/L$.  However,
it is known that this is not the
case\cite{Das:1996ug,Polchinski:1996fm,Maldacena:1996ds,Dijkgraaf:1996xw,
Hashimoto:1996pd}.
One can add a flat, but globally nontrivial connection which has
the effect of changing the boundary conditions on the fields. 
Rather than having 
$N^2$ fields each with period $L$, one can
arrange to have $N$ 
fields which are
periodic after
transversing one circle $N$ times.
The lowest energy states thus
have energy of order $1/(NL)$.  In string theory, this
construction has a simple interpretation.  The $U(N)$ gauge
theory describes the low energy excitations of $N$ D-branes
wrapped on $\T^m$.  However, it also describes the low energy
excitations of one D-brane wrapped $N$ times around a circle.
The open strings
$(i,j)$ connecting the $i^{th}$ and $j^{th}$ brane in the first case
are still present in the second, but after going around the
circle they are now identified with the $(i+1,j+1)$ strings. So 
for the wrapped brane, there are
only $N$ different types of open strings labelled by $|i-j|$
on a circle of length $NL$.
The latter configuration thus has lower energy states and corresponds
to adding the flat connection in the gauge theory.

It is natural to ask if this same construction works for other
spaces.  Consider a $U(N)$ gauge theory on the quotient space
$M/\G$ where $\G$ is any freely acting discrete group of
isometries 
of a compact Riemannian manifold $M$.  ($\G$ can be nonabelian.) 
Since the size of $M/\G$ can be much smaller than the original
space $M$, one might have thought that the energy of low lying
states is increased.  However, by analogy with the case of the
torus, one can ask if there exists a flat but globally nontrivial
connection such that the energy of states in the presence of this
connection is the same as the theory on $M$ without the quotient. 
We will show that when $N = n |\G|$ (where $|\G|$ is the number
of elements in $\G$), the answer is yes.  The condition on $N$ is
exactly what one would expect from wrapping branes.  $U(N)$ gauge
theory on $M/\G$ can be obtained by wrapping $N$ branes on
$M/\G$, but it can also be obtained by wrapping $N/|\G|$ branes
on $M$ and then taking the quotient.  In the first case each
brane is wrapped once.  In the second, each brane is wrapped
$|\G|$ times.

Although we have mentioned the energy of states at weak coupling,
the result we establish is actually more general. 
Nontrivial flat connections correspond to different
classical vacua. Since our space is compact,  one must sum over these
vacua in the functional integral.  At low temperature, the partition 
function (even at strong coupling)
will be dominated by the sector with the most light states. This
is near the connection which makes the space look ``as big as possible".

Since one often considers branes wrapped on cycles in string
theory, this result may have various applications.  We consider
here its application to the AdS/CFT correspondence
\cite{Maldacena:1998re,Aharony:2000ti}.  In the simplest context,
this states that string theory on spacetimes which asymptotically
approach $AdS_5\times S^5$ is equivalent to a conformal field
theory (CFT), namely an ${\cal N} = 4$ super Yang-Mills theory,
on $S^3 \times R$.  The gauge group of the Yang-Mills theory is
$U(N)$, where $N$ is related to the string coupling $g$, string
length $\ell_s$, and the AdS radius $R$, by $N=(4\pi
g)^{-1}(R/\ell_s)^4$.  The Yang-Mills coupling constant is
$g^2_{\rm YM}=4\pi g$.  In order for the asymptotic time
translation in $AdS_5$ to agree with Hamiltonian evolution in the
gauge theory, one must take the radius of the $S^3$ to be the
same as the radius of the $AdS$ spacetime.  This follows from the
fact that if one rescales the AdS metric
\beq\label{ads}
ds^2 = - \( {r^2\over R^2} +1\) dt^2 + 
\( {r^2\over R^2} +1\)^{-1} dr^2  + r^2 d\Omega_3
\eeq
by $R^2/r^2$ so that $t$ is proper time in the asymptotic metric,
the radius of the $S^3$ is $R$. In a Hamiltonian formulation 
\cite{Witten:1998qj,Horowitz:1998bj}, the
equivalence of these two theories implies that the energies of
all states must agree.  Of course, in most cases, one cannot
compute the energy of states in the strongly coupled field
theory.  But this is possible for states protected by
supersymmetry, which includes the gauge theory analogs of all the
supergravity modes.

To obtain the dual description of the gauge theory on $S^3/\G$,
we must consider spacetimes which asymptotically approach the
quotient of $AdS_5$ by a freely acting discrete subgroup $\G$ of
$SO(4)$ acting on the three-spheres of spherical symmetry.
If we start with global
$AdS_5$, this produces an orbifold singularity at $r=0$.  We will
make some comments about this orbifold in section 5, but it is
difficult to analyze exactly due to the Ramond-Ramond background. 
However we can confirm the existence of extra light states and
avoid the orbifold singularity by going to finite temperature. 
At temperatures $T>T_{crit} \sim 1/R$, the 
thermal state of the
gauge theory is
described by the Schwarzschild-AdS solution.  One can now take
the quotient by $\G$ without producing any additional
singularities.  Using this dual description, one can ask when
finite size effects become important.  
We will see from the gravity theory
that the answer is $T\sim 1/R$ for all $\G$
(provided $|\G|<R/l_s$).  Since the
volume of $S^3/\G$ is much smaller than $S^3$, this is possible
in the gauge theory
only if there are extra light states.

In the next section, we show the existence of a nontrivial flat
connection on $M/\G$ with the desired properties.  In section 3
we review the case of gauge theory on $\T^3$ and show that the
existence of extra light states in this case can be seen from the
AdS/CFT correspondence by considering nonzero temperature.  A
similar argument allows one to infer the existence of extra light
states of the gauge theory on $S^3/\G$.  This is discussed in
section 4, and a final section contains some remarks about the zero
temperature case (where one has an orbifold of AdS). For a discussion
of bound states of branes wrapped on $M/\G$ with  
nontrivial flat connections see \cite{Gopakumar:1997dv}. 

\section{Nontrivial flat connection on  $M/\G$}  
\setcounter{equation}{0}

In this section we consider a $U(N)$ gauge theory on $M/\G$,
where $\G$ is any freely acting discrete isometry group of $M$. 
We want to find a flat connection so that the
density of modes is essentially indistinguishable from the same
theory on $M$.  More precisely, consider a scalar
field\footnote{We are actually interested in fields valued in the
hermitian matrices.  This restriction will not lead to any
complications however.} $\phi:\Mg\rightarrow C^N\otimes C^N$
transforming under the adjoint representation of $U(N)$.  (The
gauge theory also has spinors and gauge fields in the adjoint,
and the following argument will apply {\it mutatis mutandis} for
them as well.)  For clarity we will focus on the eigenmodes of
the gauge covariant Laplacian.  At weak coupling, the
corresponding eigenvalues are directly related to the energy of
states.  
Normally, one expects the eigenvalues to be
larger on $M/\G$, but we will show the following: 
\begin{quote}
\noindent There exists a flat
connection $A$ such that for every eigenvalue $\lambda$ of the
Laplacian on $M$ (with zero connection) there exist solutions to
$D_A^2\phi=\lambda \phi$ on $M/\G$.  Furthermore, the number of
such modes is $1/\oG$ times the number of corresponding modes on $M$.
\end{quote}
Since the volume of $M/\G$ is also reduced by $|\G|$, the number of
modes per unit volume is the same as on $M$.

We shall first consider the case $N=\oG$.  The argument is then
easily generalized to the case when $N=n\oG$ for some integer
$n$.  In the case where $N=|\G|$ the gauge group is $U(|\G|)$. 
The required nontrivial flat $U(|\G|)$ connection is defined as
follows.  $M\rightarrow\Mg$ is a nontrivial principal $\G$-bundle
\cite{KN}.  The regular representation ${\R}:\G\rightarrow
U(\oG)$ allows us to define an associated principal
$U(\oG)$-bundle, using the left action of ${\R}(\G)$ on $U(\oG)$. 
This associated bundle is reducible to an ${\R}(\G)$ bundle that
can be identified with the original $\G$-bundle $M$.  The
original bundle $M\rightarrow\Mg$ carries a natural flat $\G$
connection, since the fiber is discrete: the horizontal subspace
at a point $u$ in $M$ is just the full tangent space to $M$ at
$u$.  This connection induces the required connection $A$ on the
associated bundle, with holonomy group ${\R}(\G)$.

A field $\phi$ 
on $\Mg$
transforming under the adjoint action of 
the gauge group $U(|\G|)$ can be 
characterized 
as the pullback
$s^*\tphi$, by a local section $s:\Mg\rightarrow M$,
of a map $\tphi:M\rightarrow C^{\oG}\otimes C^{\oG}$
transforming as 
\beq
\tphi(u\cdot\g)= {\R}(\g^{-1})\tphi(u){\R}(\g)
\label{equivariance}
\eeq
under the right action of $\G$ on $M$.  That is, $\tphi$ is
``equivariant" under the right action of $\G$ in the adjoint of
the regular representation of $\G$.  
This global characterization of $\phi$ as the pullback of an
equivariant map $\tphi$ on $M$ encodes the effect of the flat
connection.  It is equivalent to specifying ``twisted boundary
conditions" for $\phi$ on $\Mg$ in the following sense.  If $\Mg$
is covered by a collection of horizontal local sections
$\{s_\a:U_\a\rightarrow M\}$, then in each of the local gauge
patches the connection $s_\a^* A$ vanishes, but on the overlaps
$U_\a\cap U_\b$ there is generally a nontrivial gauge
transformation relating $s_\a^* \tphi$ to $s_\b^* \tphi$. 
Because of the nontrivial holonomy of $A$ there is no global
horizontal section, so not all of these non-trivial overlaps can
be removed.  On a circle, there would be just one non-trivial
periodicity condition, while on $\s3g$ these ``twisted boundary
conditions" are generally more complicated.

Viewed in this way, the fields $\phi$ on $\Mg$ in the presence of
the connection $A$ are just the equivariant subclass of all the
fields on $M$.  For each gauge orbit $u\cdot\G$ in $M$,
$\tphi$ can be specified freely only at one point, its value at
the rest of the points being determined by the equivariance
condition (\ref{equivariance}).  Since there are $\oG$ points in
an orbit, it is plausible that the number of equivariant
$\tphi$-modes is $1/\oG$ times the number of unrestricted 
modes on $M$.  We now show that this is indeed the case.

The solutions to the mode equation $D_0^2\psi=\lambda \psi$ on
$M$ can be decomposed according to unitary irreducible
representations of the action of $\G$.  The case of a general
group will be treated below using some results from
representation theory, but it is perhaps helpful to the reader to
first treat the more elementary case where $\G$ is a cyclic group
generated by some group element $\g$.  This is basically the case
that has been previously treated in the literature
\cite{Hashimoto:1996pd} for the quotient $S^1/\G$.  The unitary
irreducible representations are then one-dimensional and
characterized by a root of unity $\exp(i2\pi k/|\G|)$:
\beq
\psi(u\cdot\g)=e^{i2\pi k/|\G|}\psi(u)
\label{psigamma}
\eeq
for some integer $k<|\G|$ (since $\g^{|\G|}=1$).  For such a
$\psi$ the equivariance condition (\ref{equivariance}) becomes a
local condition
\beq
{\R}(\g^{-1})\psi(u){\R}(\g)=e^{i2\pi k/|\G|}\psi(u)  
\label{gammaequi}
\eeq
on the matrix $\psi(u)$.  We need to determine the dimension of
the subspace of $|\G|\times|\G|$ matrices satisfying this
condition.

To this end let us use the eigen-basis $\{|p\ra\}$
of ${\R}(\g)$:
\beq
{\R}(\g)|p\ra=e^{i2\pi p/|\G|}|p\ra, \qquad p=1,\dots, |\G|.
\eeq
A basis for the space of $|\G|\times|\G|$ matrices is 
$\{|p\ra\la q|\}$.  
Under the adjoint action of ${\R}(\g)$, these basis
matrices transform as
\beq
 {\R}(\g^{-1})|p\ra\la q| {\R}(\g)=e^{-i2\pi (p+q)/|\G|} |p\ra\la q|.
\label{pqtransform}
\eeq
The matrix $|p\ra\la q|$ thus satisfies the equivariance relation
(\ref{gammaequi}) provided $p+q=-k$ modulo $|\G|$.  The number of
such $(p,q)$ pairs is $|\G|$, so there is a $|\G|$-dimensional
equivariant subspace of the the $|\G|^2$-dimensional space of
matrices.  That is, there are $1/|\G|$ times as many equivariant
modes as there are unconstrained modes, which is what we set out
to prove.

In the above discussion we worked with all the matrices
in $C^{\oG}\otimes C^{\oG}$, but we are really interested in
fields taking values in the hermitian matrices, for which 
the set $\{|p\ra\la q|+|q\ra\la p|\}$ forms a basis. 
Since the transformation rule (\ref{pqtransform}) is symmetric
in $p$ and $q$, these basis matrices transform the same
way as the non-hermitian ones, so the previous analysis 
restricts without modification to the hermitian subspace.

Now let us generalize to an arbitrary non-abelian subgroup $\G$. 
In this case the irreducible representations are no longer
one-dimensional, but fortunately representation theory easily
yields the result as follows.  To begin with let us reexpress the
equivariance condition (\ref{equivariance}) as
\beq
{\R}(\g)\tphi(u\cdot\g){\R}(\g^{-1})= \tphi(u), 
\label{singlet}
\eeq
that is, the matrix-valued function must transform under the
trivial representation of $\G$ when both the argument (the point on
$M$) and the matrix are simultaneously transformed.  The
counting problem then reduces to this: how many singlets are
there among the matrix-valued functions on $M$?

The matrix valued functions carry the representation ${\cal
S}\otimes({\R}\otimes {\R}^*)$, where $\cal S$ is the vector
space of scalar-valued functions and here $\R$ stands for the
vector space of the regular representation and $\R^*\simeq\R$ is
its dual.  Let ${\W}$ be one of the irreducible representations
into which $\cal S$ decomposes, and
consider the matrix-valued functions transforming under ${\W}$ in
the argument.  These carry the representation
$\W\otimes({\R}\otimes {\R})$, where $\R$ has now been identified
with its dual.

To find the number of singlets we need three results from
representation theory \cite{Serre}. First, all irreducible
representations of a finite group are finite dimensional.
Second, the singlet occurs once in $\R$. Third, if ${\V}$ is any irreducible
representation then ${\V}\otimes {\R} = (\mbox{dim} {\V}){\R}$.
The first two results follow from 
the fact that each irreducible representation occurs in $\R$ with multiplicity
equal to its dimension.  The third result can be seen easily
using the multiplicative property of characters:
$\chi_{\V\otimes \R}=\chi_\V \cdot \chi_\R$.  The character
of the regular representation vanishes everywhere except at the
identity, so the same is true for $\V\otimes \R$.  At the
identity we have $\chi_{\V\otimes
\R}(e)=\chi_\V(e)\cdot\chi_\R(e)=({\rm dim}\V)|\G|$, hence
$\chi_{\V\otimes \R}= \chi_{({\rm dim}\V)\R}$.  Since the
character determines the representation, it follows that
$\V\otimes \R= (\mbox{dim} {\V})\R$.

Using the 
first and third results,
we have $\W\otimes({\R}\otimes
{\R})=(|\G|{\rm dim}\W)\R$.  Using the second result, there are
therefore $|\G|{\rm dim}\W$ singlets.  This number is indeed
$1/|\G|$ times the dimension of the full space of matrix-valued
functions transforming in the argument via ${\W}$, namely
$|\G|^2{\rm dim}\W$.

The counting is now easily generalized to the case where
$N=n|\G|$ for some integer $n$.  
As in the case $N=|\G|$, there is a principal $U(n|\G|)$-bundle
associated to the $\G$-bundle $M\rightarrow\Mg$, now
by the left action of $\R(\G)$ on $U(n|\G|)$ via
the $n$-fold direct sum $n{\R}(\G)$ of the regular representation.
As before, this associated bundle is reducible to a 
$\G$-bundle equivalent to $M\rightarrow\Mg$, and 
the nontrivial flat $\G$-connection on $M\rightarrow\Mg$
induces the required flat connection on the $U(n|\G|)$-bundle.
The fields $\phi:M/\G\rightarrow C^{n|\G|}\otimes C^{n|\G|}$ 
transforming
under the adjoint representation of $U(n|\G|)$ are now realized
as maps $\tphi:M\rightarrow C^{n|\G|}\otimes C^{n|\G|}$ that are
equivariant under the representation $n{\R}(\G)\subset U(n|\G|)$ 
of $\G$, i.e. as before they are singlets under the combined action
of $\G$ on the argument and the matrix value. 
The $N\times N$ matrices carry the representation
$n{\R}\otimes n{\R}= (n^2|\G|){\R}$, and the matrix-valued
functions transforming under ${\W}$ in the argument carry
$(n^2|\G|){\W}\otimes {\R}= (n^2|\G|{\rm dim}\W) \R$.  The
dimension of the space of singlet functions is
therefore $n^2|\G|\mbox{dim} {\W}$, which is
again $1/|\G|$ times the dimension of the full space of $N\times
N$ matrix-valued functions transforming in the argument via
${\W}$.

We have so far shown that the nontrivial flat connection $A$ 
makes the space $\Mg$ ``look bigger" for fields valued in 
the adjoint of $U(N)$. The argument can also be applied to 
fields valued in the fundamental representation.  These are
maps $\psi:\Mg\rightarrow C^N$, which can be characterized
as local pullbacks of an equivariant map 
$\widetilde{\psi}:M\rightarrow C^N$. 
Let us consider here the case $N=|\G|$. The generalization
to $N=n|\G|$ follows that given above for fields in the
adjoint. 
The equivariance condition is then 
$\widetilde{\psi}(u\cdot \g) =\R(\g^{-1})\widetilde{\psi}(u)$,
which implies that $\widetilde{\psi}$ is a singlet in ${\cal S}\otimes 
\R$. Those $\widetilde{\psi}$ transforming in the 
argument under the irreducible representation $\W$ carry the representation
$\W\otimes \R= ({\rm dim} \W)\R$.  This contains $({\rm dim} 
\W)$ singlets, which is again $1/|\G|$ times the dimension of 
the full space of fields in the fundamental transforming under 
$\W$ in the argument.

\sect{Gauge theory on $\T^3$ at nonzero temperature}
\setcounter{equation}{0}

We now turn to applications of the above result using the AdS/CFT
correspondence.
Before discussing the case of gauge theory on $S^3/\G$,
we consider the simpler case of gauge
theory on $\T^3$.  The dual supergravity description of ${\cal N}
=4$ super Yang-Mills at temperature $T$ on a three torus is given
by the near horizon limit of a near extremal black three brane. This is
the product of $S^5$ and 
\beq\label{threeb}
ds^2 = {r^2 \over R^2} 
\[ -\(1-{r_0^4\over r^4}\) dt^2 + dx_idx^i \]
+ \(1-{r_0^4\over r^4}\)^{-1} {R^2 dr^2 \over r^2}
\eeq
where the three coordinates $x_i$ are periodically identified.
The Hawking temperature is $T=r_0/(\pi R^2)$ and the total mass 
is
\beq\label{mtorus}
M = {3\pi^2\over 8} N^2T^4 V_3
\eeq
where $V_3$ is the volume of the three torus and we have used the
fact that $G_5= \pi R^3/2N^2$.  This is known to agree with the
weakly coupled gauge theory up to an overall factor of
$3/4$ \cite{Gubser:1996de}.

If the coordinates $x_i$ each have length $L$, one would expect
finite size effects to show up in the gauge theory when $T \lsim
1/L$.  However as long as the metric (\ref{threeb}) is valid,
supergravity predicts (\ref{mtorus}) with no finite size
corrections. 
The supergravity solution is expected to break down
when either the curvature at the horizon or the proper lengths of
the compact directions at the horizon are of order the string
scale $\ell_s$ .  The curvature at the horizon is always of order
$1/R^2$, so the curvature is never a problem as long as the AdS
radius is much larger than the string scale, $R\gg\ell_s$.  The
second condition is satisfied when $r_0 L/R\sim\ell_s$ or
$TLR\sim \ell_s$.  So when $TL \sim 1$, the compact directions at
the horizon are still much larger than the string scale, and
there should be no corrections.  This means that super Yang-Mills
on $\T^3$ must have states lighter than $1/L$ so the finite size
corrections are suppressed.

As discussed in the introduction, it has already been shown in other contexts 
that this is indeed
the case.
Since the potential energy is positive definite, 
the energy of each state should increase as the coupling increases\footnote{One
can show that the potential energy is of the same order as the kinetic
energy \cite{Horowitz:1996nw}, so the energy of each state increases by only  
a factor of two or so. This provides a qualitative explanation of the factor 
of $3/4$ discrepancy between the weak and strong coupling results.}.
So evidence for light states at strong coupling implies that there must
be light states at weak coupling.
These light states arise by introducing a flat, but nontrivial,
connection on $\T^3$.
If one views the gauge theory as
describing the low energy excitations of 
3-branes,
the
standard description without a background connection corresponds
to $N$ branes each wrapped once around the torus.  The flat
connection describes multiply wrapped branes.  If one brane is
wrapped $N$ times around one circle, there will be very light
states with energies of order $1/(NL)$, but the effective theory
of these states will be one dimensional, and their contribution
to the energy will be relatively small.  To maintain a three
dimensional theory (as indicated by (\ref{mtorus})) one considers
instead a single brane wrapped $N^{1/3}$ times around each of the
three circles.  Then the excitations have energy in multiples of
$1/(N^{1/3}L)$, so we would not expect to see finite size
corrections until $TL\sim N^{-1/3}$.  We have seen that string
theory allows finite size corrections to supergravity when $TL
\sim
\ell_s/R = (4\pi gN)^{-1/4}$.  Since $(gN)^{1/4} \ll N^{1/3}$
(for large $N$ and $g<1$) one sees that 
finite size effects in the weakly coupled gauge theory do not become important
until a temperature well below that at which they are allowed by
the correspondence with string theory. 
One might even conclude that the string winding mode corrections on the 
gravity side should not affect the relation between energy and temperature
until $T\sim 1/(N^{1/3}L)$. On the other hand, it might be that at strong
coupling in the gauge theory, finite size effects already become important
at the higher temperature $T\sim (4\pi gN)^{-1/4}L^{-1}$.

\section{Gauge theory on $S^3/\G$ at nonzero temperature}
\setcounter{equation}{0}

We turn now to the case of a gauge theory on $S^3/\G$.  Let us
first ask how one sees finite size effects due to the $S^3$ (with
no quotient) using the AdS/CFT correspondence.  A thermal state
in the gauge theory at high temperature is described on the
supergravity side by a large Schwarzschild AdS black hole
\beq\label{sads}
ds^2 = -f(r) dt^2 + f^{-1}(r) dr^2 + r^2 d\Omega_3
\eeq
with 
\beq
f(r) = {r^2\over R^2} +1- {r_0^2\over r^2}
\eeq
The Schwarzschild radius $r_+$ is defined by $f(r_+)=0$.  The ADM
mass (neglecting the constant $AdS_5$ contribution discussed in the next 
section) is
\beq
M={3\pi r_0^2\over 8 G_5} 
\eeq
where $G_5$ is the five dimensional Newton's constant.  The
Hawking temperature is\footnote{Given $T$, there are two
solutions to this equation for the Schwarzschild radius $r_+$. 
We are interested in the larger one, $r_+>R$, since small black
holes have negative specific heat.}
\beq\label{temp}
T={R^2 + 2r_+^2\over 2\pi r_+ R^2}
\eeq
It follows that black holes have a minimum temperature
$T_{min}=\sqrt 2/\pi R$.  
At low temperature, the corresponding
supergravity solution is just a gas of particles in $AdS_5$.  The
temperature at which the description changes can be computed by
comparing the action for the euclidean black hole and $AdS_5$
with imaginary time periodically identified.  This was done in
\cite{Witten:1998zw} (following \cite{Hawking:1983dh}) with the
result that the black hole dominates for $r_+>R$ corresponding to
$T>T_{crit}= 3/(2\pi R)$.  We assume this is the case in the
following.

We wish to compute how the mass changes as a function of
temperature.  (One could equally well compute the entropy as a
function of temperature, but that will be determined in terms of
the mass since $dM = T dS$.)  For convenience, we set $R=1$ and
measure all quantities in units of the AdS radius.  First note
that since $f(r_+)=0$,
\beq\label{rzero}
r_0^2 = r_+^2 + r_+^4
\eeq
Using (\ref{temp}) we solve for $r_+$ in terms of $T$:
\beq\label{rtrel}
2 r_+ = \pi T + \pi T \(1 -{2\over \pi^2 T^2}\)^{1/2}
\eeq
Substituting this into (\ref{rzero}) yields
\beq
r_0^2 = {1\over 2} \pi^4 T^4 \[1+\(1-{2\over \pi^2T^2}\)^{1/2}\] 
   -{\pi^2 T^2\over 2} - {1\over 4}
\eeq
To leading order for large temperature, the mass is
therefore\footnote{This calculation of the supergravity energy at
given temperature is actually an underestimate, since we have
neglected the energy in the Hawking radiation outside the black
hole.  However, as long as the AdS radius is large compared to
the Planck scale, this contribution is negligible.}
\beq\label{sanw}
M= {3\pi\over 8G_5} [ \pi^4 T^4 - \pi^2 T^2 + O(T^0) ]
\eeq

We now want to compare this with the energy of a weakly coupled
$U(N)$ SYM. This calculation was done in 
\cite{Burgess:1999vb,Hawking:1999dp}.
It was found that finite size effects
alter the factor of 3/4 that relates the energies in the high
temperature limit.  Here we recap the result.

We first translate the supergravity result (\ref{sanw}) into
field theory language.  Since the volume of a unit three sphere
is $V = 2 \pi^2$ and $G_5 = \pi/2N^2$, this becomes
\beq\label{sugrae}
M = {3 N^2 V\over 8} (\pi^2 T^4 -T^2 )
\eeq
The first term agrees with the torus result (\ref{sugrae}).  The
second corresponds to finite size effects in the gauge theory. 
It is important only when $T$ is of order one in units of the AdS
radius, i.e, $T\sim 1/R$.

To compute the weakly coupled gauge theory energy directly, we
need to calculate the modes for scalars, spinors and vectors on
$S^3$, and evaluate the sums which yield the average energy at a
given temperature.  One way to do so is to use the results of
this calculation from the ancient Ref.  \cite{AltaieDowker}.  The
result is that for large temperature, the energy density for a
single scalar, Weyl fermion, or vector is
\beq
e_0 = {1\over 30} \pi^2 T^4, \qquad e_{1/2} 
={7\over 120} \pi^2 T^4 -{T^2\over 48}, 
\qquad e_1 = {1\over 15} \pi^2 T^4 - {T^2\over 6}
\eeq
The first term in each case is the usual blackbody expression,
and the second term is the leading finite size correction.  (For
the scalar field, there is no correction of order $T^2$.)  The
total energy for the weakly coupled SYM theory, which has six
scalars, four Weyl fermions, and one vector, is therefore $ E =
N^2 V(6e_0 + 4 e_{1/2} + e_1) $, or
\beq\label{syme}
   E= {N^2 V\over 2}\(\pi^2 T^4 - {T^2 \over 2} \)
\eeq

Comparing (\ref{sugrae}) and (\ref{syme}) we see that the leading
terms show the well known factor of $3/4$ difference between the
weakly coupled gauge theory and the supergravity prediction for
the strongly coupled regime.  The next terms show that this
factor of $3/4$ discrepancy is not preserved as we lower the
temperature.  For a given temperature, the supergravity energy
starts out smaller than the weakly coupled field theory energy,
and decreases more quickly as we lower the
temperature.\footnote{At the minimum black hole temperature,
numerical evaluation of the mode sums shows that the ratio of
black hole mass to the weakly coupled SYM energy is $\sim 0.17$.}
This implies that the interactions, which are omitted in the
weakly coupled gauge theory calculations, must become relatively
more important when the temperature is lowered.  This makes sense
since, at sufficiently low temperature, there is a critical point
at which the gauge theory becomes confining.  This critical point
corresponds to the transition where the entropy drops
precipitously and the black hole goes away.

The Schwarzschild-AdS metric (\ref{sads}) has spatial slices with
topology $S^3\times R$.  Hence one can take the quotient of $S^3$
by any freely acting discrete symmetry group $\G\subset SO(4)$
without producing additional singularities.  The temperature is
unaffected by the quotient.  The only effect on the mass
(\ref{sugrae}) is the reduction of the volume $V$ by a factor
$\G$.  Since the energy density is unchanged, the finite size
effects do not arise until $T\sim 1/R$ as before.  Similarly, in
comparing the euclidean action for Schwarzschild-AdS metric and
periodically identified AdS, the quotient affects both of them
equally\footnote{The orbifold singularity does not contribute to the 
gravitational action since if one removes a small tubular neighborhood 
of the singularity with radius $\epsilon$, the extra surface term 
at this inner boundary goes to zero as $\epsilon \rightarrow 0$.}
by changing an overall factor corresponding to the volume
of a unit $S^3$. 
Thus the critical temperature at which the
black hole geometry dominates is also unchanged.  The
supergravity solution can be trusted as long as string
corrections are not important.  Since $r_+ >R$ and $R\gg \ell_s$,
the Schwarzschild radius is much larger than the string
scale.  The other condition to check is that all noncontractible
loops have size much larger than $\ell_s$.  This will be true as
long as $\oG \ll R/\ell_s$. (This condition is sufficient, but not necessary.
Depending on how $\G$ acts on the sphere, we could have $\oG$ larger
than $R/\ell_s$ -- but certainly no larger than $(R/\ell_s)^3$ --
and still have no noncontractible curve with length less than $\ell_s$.)
Now consider the gauge theory side. 
Even though the volume of space is drastically reduced (from
$S^3$ to $S^3/\G$), the AdS/CFT correspondence implies that 
the density of low energy states is not affected.  
This is possible only if one adds the nontrivial flat connection
discussed in section 2.

The result in section 2 assumed that $N/\oG$ is an integer.  This
was needed in order to construct a connection with holonomy in
the direct sum of $N/\oG$ copies of the regular representation of
$\G$.  In terms of D-branes, this corresponds to assuming that
each brane is wrapped $\G$ times, yielding a total of $N$ local
branes.  If $N/\oG$ is not an integer then the result cannot hold
precisely, and one would expect the relative size of the
corrections will be of the order of $\oG/N$.  These corrections
will be small unless $\oG$ is comparable to $N$.  As noted above,
however, validity of the supergravity solution outside the black
hole horizon at finite temperature requires that the length of
any noncontractible curves be much longer than $l_s$.  This
implies that $\oG < (R/l_s)^3=(4\pi g N)^{3/4}\ll N$.  Hence string
corrections to supergravity are already expected before $\G$
becomes large enough for non-integer $N/\oG$ to be an issue.

\section{Gauge theory on $S^3/\G$ at zero temperature}
\setcounter{equation}{0}

We have discussed nonzero temperature above, but it is natural to
ask whether one can also see these extra light states at zero
temperature.  In this case, the spacetime is just $AdS_5$, and we
want to consider the quotient by a discrete subgroup $\G$ of
$SO(4)$ acting on the three-spheres of spherical symmetry.  This
produces an orbifold with fixed points at $r=0$.  Starting with
\cite{Kachru:1998ys}, there has been considerable interest in the
effect on the AdS/CFT correspondence of taking quotients
$S^5/\G$.  These orbifolds reduce the amount of supersymmetry and
allow new phenomena.  There has been relatively less
investigation of orbifolds of $AdS_5$ (although see e.g.
\cite{Ghosh:1999nf,Martinec:2001cf,Son:2001qm}).
It is difficult to calculate
the precise spectrum of twisted sector states due to the
Ramond-Ramond background. However the following argument suggests
that the extra light states in the gauge theory are necessary for
consistency of the AdS/CFT correspondence in this case also.

Consider the low energy excitations on the gravity side.  There
are the usual supergravity modes which are invariant under $\G$. 
Before the quotient, these modes correspond to states on $S^3$
which are invariant under $\G$.  So they exist in the theory on
$S^3/\G$ without an extra flat connection.  However, there are
also modes in the twisted sector localized in AdS/$\G$ 
near the fixed point.  These should be similar to the modes in
the twisted sector of a flat space orbifold.  If $\G$ breaks all
the supersymmetry, then there are tachyons
\cite{Adams:2001sv} representing an
instability in the strongly coupled gauge theory.  
It was shown
in \cite{Ghosh:1999nf} that if $\G$ lies in an $SU(2)$ subgroup
of $SO(4)$, then half the AdS supersymmetry is preserved.  (This
is the same condition for orbifolds in flat space.)  In this
case, it is plausible that there are massless modes.  
These modes are localized near the origin of $AdS_5$, so they
correspond to fields on $S^5$ cross time.  Since the $S^5$ has
radius $R$ and is unaffected by the quotient, these fields
produce states with energy of order $1/R$.  In the gauge theory,
these light states can be reproduced only with the nontrivial
flat connection.

A less direct way to see the existence of the extra light states
is by looking at the ground state energy density. A boundary stress-energy
tensor for asymptotically AdS spacetimes was proposed in 
\cite{Balasubramanian:1999re}. For $AdS_5$,
this gives a finite nonzero answer for the total energy\footnote{It was
pointed out in \cite{Ashtekar:1999jx}
that one cannot assign a nonzero energy to pure AdS
without breaking the symmetry. However, as explained in \cite{Skenderis:2000in},
the  AdS symmetry is indeed broken by the
need to regulate infrared divergences.}
$E=3\pi R^2/32 G_5$. 
This turns out to agree
exactly with the Casimir energy of ${\cal N} =4$, $U(N)$ super Yang-Mills on
a three-sphere of radius $R$. (The Casimir energy can be computed even
at strong coupling since it is determined by the trace anomaly which,
due to supersymmetry, does not receive corrections.) 
Now imagine taking the quotient
by $\G$. On the gravity side,  since the local geometry is unchanged and the
volume of spheres are reduced, it is clear that the total energy will be
reduced by a factor of $|\G|$. If this is identified with 
the ground state energy
of the gauge theory on $S^3/\G$, then the energy density is unchanged by
the quotient. This is possible only due to the nontrivial flat connection.

Before one can conclude that the  AdS/CFT correspondence
predicts that the ground state
energy density of the gauge theory on  $S^3/\G$ is the same as on $S^3$,
one needs to answer the following open question:
Do there exist solutions of Einstein's equation
with negative cosmological constant which approach $AdS_5/\G$
asymptotically and have less energy than  $AdS_5/\G$?
If so, AdS/CFT would predict that the ground state energy density of the 
strongly coupled gauge theory on $S^3/\G$ is even smaller than
that of the gauge theory on $S^3$. Although this seems unlikely,
it has been shown that with zero cosmological constant, 
there are smooth $4+1$ dimensional vacuum solutions
which are asymptotically locally euclidean and have negative
total energy \cite{LeBrun}.  In fact the energy is unbounded from below. 
In the case of Kaluza-Klein boundary conditions (one direction compactified
on a circle), it is also true that there are asymptotically flat vacuum 
solutions with arbitrarily negative energy. However, in this case, there is
evidence that when the cosmological constant is negative, the energy is bounded
from below \cite{Horowitz:1998ha}. So if there exist solutions asymptotic to 
$AdS_5/\G$ with energy less than $AdS_5/\G$, it is plausible that
the energy will at least be bounded from below.

\section*{Acknowledgements}
We thank J. Rosenberg for instructive discussions on 
representation theory. The work of G.H. was supported in part by
NSF grant PHY-0070895. 
The work of T.J. was supported in part by the NSF under grant
PHY-9907949 at the ITP in Santa Barbara and grant PHY-9800967
at the University of Maryland.

\end{document}